\documentstyle[11pt,amssym,aasms4]{article}

\begin{document}


\title{THE MASS-TO-LIGHT FUNCTION: \ ANTIBIAS AND $\Omega _{m}$}
\author{N.A. Bahcall, R. Cen, R. Dav\'{e}, J.P. Ostriker, Q. Yu}
\affil{Princeton University Observatory, Princeton, NJ 08544}

\begin{abstract}

We use large-scale cosmological simulations to estimate the
mass-to-light ratio of galaxy systems as a function of scale, and
compare the results with observations of galaxies, groups, clusters, and
superclusters of galaxies. We find remarkably good agreement between
observations and simulations. Specifically, we find that the simulated
mass-to-light ratio increases with scale on small scales and flattens to a
constant value on large scales, as suggested by observations. We find that
while mass typically follows light on large scales, high overdensity regions
--- such as rich clusters and superclusters of galaxies --- exhibit higher
$M/L_{\rm B}$ values than average, while low density regions exhibit lower
$M/L_{\rm B}$ values; high density regions are thus \emph{antibiased} in 
$M/L_{\rm B}$, with mass more strongly concentrated than blue light. This
is true despite the fact that the galaxy mass density is unbiased or
positively biased relative to the total mass density in these regions.
The $M/L_{\rm B}$ antibias is likely due to the relatively old age of the high 
density regions, where light has declined significantly since their early 
formation time, especially in the blue band which traces recent star formation.
Comparing the simulated results with observations, we place a powerful
constraint on the mass density of the universe; using, for the first time,
the entire observed mass-to-light function, from galaxies to superclusters,
we find $\Omega =0.16\pm0.05$.

\end{abstract}

\section{ INTRODUCTION}
One of the oldest - and simplest - techniques for estimating the mass
density of the universe is the mass-to-light method. In this method, the
average ratio of the observed mass to light of the largest possible systems
is used; assuming it is a fair sample, it can then be multiplied by the
total luminosity density of the universe to yield the universal mass
density. When the method is applied to rich clusters of galaxies --- the
largest virilized systems for which a mass has been reliably determined --- the
total mass density of the universe adds up to only $\Omega \simeq 0.2\,\ $%
(where $\Omega $ is the mass density in units of the critical density)
(Zwicky 1957, Abell 1965, Ostriker, Peebles \& Yahil 1974, Bahcall 1977,
Faber \& Gallagher 1979, Trimble 1987, Peebles 1993,
Bahcall, Lubin \& Dorman 1995, Carlberg \emph{et al.} 1996, 1997, and
references therein). A fundamental assumption in this determination,
however, is that the mass-to-light ratio ($M/L$) of clusters is a fair
representation of the universal value. If the mass-to-light ratio of
clusters is larger or smaller than the universal mean, then the resulting $%
\Omega $ will be an over- or under- estimate, respectively. It is
not clear whether this classic assumption of an unbiased
representation by clusters is correct. More generally, if mass follows
light (i.e., galaxies) on large scales --- thus $M/L\simeq $ constant --- the
galaxy distribution is considered to be unbiased with respect to mass; if
mass is distributed more broadly than light, as is generally believed, then
the galaxy distribution is biased (i.e., more clustered) with respect to
mass, and the above determination of $\Omega $ is an underestimate. \ We
investigate these questions of cluster representation and bias, and the
impact they have on the measurement of $\Omega $.

Observations of galaxies, groups and clusters of galaxies suggest that 
$M/L$ increases as a function of scale up to scales of hundreds
of kiloparsecs (Schwarzschild 1954, Rubin \& Ford 1970, Roberts \& Rots 1973,
Ostriker \emph{et al.} 1974, Einasto \emph{et al.} 1974, Davis \emph{et al.}
1980, Trimble 1987, Gramann 1990, Zaritzky \emph{et al.} 1993, Fischer \emph{et al }$1999$),
but then flattens out and remains approximately constant on larger scales
(Bahcall, Lubin \& Dorman 1995). In the modern context we normally interpret
this fact as indicating that luminous galaxies are more concentrated in peak
density regions than the dark matter because baryons are dissipational. The
shape and amplitude of the mass-to-light function --- that is, the dependence
of $M/L$ on scale, $(M/L)(R)$ --- can place powerful constraints on
the amount and distribution of dark matter in the universe, as well as on
the amount of bias and its dependence on scale. The $M/L$ function thus provides
a direct, model-independent census of the total mass density of the universe.

What is the expected dependence of $M/L$ on scale? In this paper we
investigate this question  using large-scale, high resolution hydrodynamic
cosmological simulations that contain dark matter and gas, and compare the
results with observations. We find an excellent agreement between models
and observations in the shape of the $M/L$ function; both data and models show
an increase on small scales (hundreds of kpc) and a flattened $(M/L)(R)$
distribution on large scales. We use the comparison between data and
simulations to determine the mass density of the universe. The amount of
bias and its dependence on scale are also revealed. \ We find that clusters
of galaxies are mildly \emph{antibiased, } in the sense that mass is more
concentrated than light on average. Previous determinations of $\Omega$
using clusters of galaxies have thus \emph{overestimated} $\Omega $ due
to this unaccounted antibias. The present investigation attempts to provide
an unbiased determination of $\Omega$
using, for the first time, the entire observed mass-to-light function.
The above results do not disagree with previous estimates that the mass density of galaxies is unbiased or positively biased with respect to
the total mass density in the high
density regions; it is the light density that is shown here to be antibiased.

\section{ OBSERVATIONS}
The observed mass-to-light ratio of galaxies, groups and clusters as a
function of scale, $(M/L_{\rm B})(R)$, is taken from Bahcall, Lubin and Dorman
(1995, hereafter BLD). In these data, masses are determined using different
methods including velocity dispersion, gravitational lensing, and X-ray gas
temperature. The luminosity L$_{B}$ throughout this paper refers to the
\emph{total } blue luminosity, corrected for both Galactic and internal
extinction. The data for rich clusters (at $R=1.5h^{-1}$Mpc) and for groups
($R\simeq 20$kpc to $1h^{-1}$Mpc), shown in the figures below, represent
median $M/L_{\rm B}$ values of large samples, as does the $M/L_{\rm B}$ ratio
for the luminous parts of typical $L_*$ elliptical and spiral galaxies
(see BLD for details). More recent observations of rich clusters from the
CNOC cluster survey (Carlberg \emph{et al. 1996, 1997}) yield consistent
results. Based on the available data, BLD find that the $M/L_{\rm B}$ ratio
of galaxy systems increases linearly with scale up to the scale of very
large galactic halos ($R\sim 0.2h^{-1}$Mpc), but then flattens on larger
scales; they suggest that $M/L_{\rm B}$ does not increase significantly with
scale beyond $\sim0.2h^{-1}$Mpc. Furthermore, BLD show that $M/L_{\rm B}$ of
elliptical galaxies is approximately three times larger than that of spirals
(at the same radius); both increase linearly with scale up to 
$R\sim0.2h^{-1}$Mpc. The total mass of groups and clusters can then be
accounted for by the combined mass of their elliptical and spiral galaxy
members, including their large halos, plus the intracluster gas. The large
halos are likely to be stripped off in the dense environments of clusters,
but their mass still remains in the clusters. Observations of weak
gravitational lensing by foreground galaxies using the Sloan Digital Sky
Survey (Fischer \emph{et al}. 1999) find consistent results indicating large
halos around galaxies.

Recently, the first determination of the mass and mass-to-light ratio of a
large supercluster ($\sim6h^{-1}$Mpc), MS0302, was obtained using
weak gravitational lensing (Kaiser \emph{et al.}1999). The mass and 
$M/L_{\rm B}$ ratio of three individual clusters in the supercuster as well
as the mass and $M/L_{\rm B}$ of the supercluster itself were all determined
from the weak lensing observations. The results show, quite remarkably, the
\emph{ same }$M/L_{\rm B}$ ratio for both the individual clusters and the
large supercluster ($260\pm40$ and $280\pm40h\ M_{\odot }/L_{\odot }$,
respectively) thus directly confirming a flat $(M/L_{\rm B})(R)$ function on
large scales, as suggested by BLD. This new supercluster result is added
in the figures below (converted to our standard L$_{B}$ system by adding the 
30\% contribution to the luminosity from spiral galaxies 
(Kaiser \emph{et al.} 1999), correcting for passive luminosity evolution from
$z=0.42$ to $z\simeq0$ following $L_{\rm B}\propto(1+z)$, and correcting for
internal extinction ($\sim$10\%; BLD); the net correction factor is 1$\pm$0.2). 
We also show (for illustration only) the $M/L_{\rm B}$ ratio determined from
the Least Action Method at $30h^{-1}$Mpc by Tully, Shaya and Peebles (1994)
and the observed range of Virgo Infall measurements (see BLD). While these
provide less direct measures of mass than the supercluster weak lensing
result (and are thus not included in our fits), they are all consistent with
each other and with the observed flattening of $M/L_{\rm B}$ with scale. The
data are presented in the figures below. 

\section{ SIMULATIONS}

 We investigate the expected behavior of $M/L_{\rm B}$ as a function of
scale using two sets of cosmological simulations which include both dark 
matter and gas: a large-scale, $100h^{-1}$Mpc box simulation to study the
large-scale behavior of $M/L_{\rm B}$, and a smaller, higher resolution
simulation with a box size of $11.1h^{-1}$Mpc, to investigate smaller scales.
The large-scale hydrodynamic simulation, described by Cen and Ostriker (1999),
uses the shock-capturing Total Variation Diminishing method on a Cartesian
grid for gas dynamics (Ryu \emph{et al, } 1993). A Particle-Mesh (PM) code
is used for dark matter particles. An FFT is used to solve Poisson's
equation. In addition, the code accounts for cooling processes including
metal cooling and heating and incorporates a heuristic galaxy formation scheme
described by Cen and Ostriker (1999) (see also below). The cosmological model
used is a flat Cold-Dark-Matter (CDM) model, with mass density $\Omega =0.37,$
cosmologial constant density $\Omega _{\Lambda }=0.63,$ baryon density
$\Omega_{\rm b}=0.039,$ and a Hubble constant $h=0.7$ (where $H_{0}=100h$
kms s$^{-1}$Mpc$^{-1}$). A power-spectrum slope of $n= 0.95$ and
normalization $\sigma _{8}=0.8$ (the mass rms fluctuations on $8h^{-1}$Mpc
scale at $z=0$ ) were used, consistent with the cluster abundance normalization
and the COBE microwave background
fluctuations (White \emph{et al.} 1993, Ostriker and Steinhardt 1995,
Bahcall and Fan 1998). This model fits well current observational data
(e.g., Ostriker and Steinhardt 1995, Krauss \emph{et al.} 1995, Bahcall 
\emph{et al.} 1999). A periodic box of $100h^{-1}$Mpc on a side is used, 
with $512^{3}$ fluid cells and $256^{3}$ dark matter particles. The dark
matter mass resolution is $6\times10^{9}h^{-1}M_{\odot }$ and the grid cell
size is $0.2h^{-1}$Mpc. We consider only scales with radii $R \gtrsim
1h^{-1}$Mpc in this simulation, which is considerably larger than the cell
size; on these scales, the relevant gravitational and hydrodynamical physics
are accurately computed. On smaller scales we use the smaller,
higher-resolution simulation described below.

Galaxies are ``identified'' in the simulation by the procedure described in
Cen and Ostriker (1999): if a cell's mass is higher than the Jean's mass, and
if the cooling time of the gas in it is shorter than its dynamical time, 
and if the flow around the cell is
converging, then it will have stars forming inside that cell. \ The code
turns the baryonic fluid component into collisionless stellar particles
(``galaxy particles'') at a rate proportional to $m_{b}/t_{\rm dyn}$, where
$m_{\rm b}$ is the mass of gas in the cell and $t_{\rm dyn}$ is the local
dynamical time. These galaxy particles subsequently contribute to metal
production, SN energy feedback and the background ionizing UV radiation. 
This algorithm is essentially the same as in Cen and Ostriker (1992) and also
used by Katz, Hernquist and Weinberg (1996), Gnedin (1996) and Steinmetz
(1996). The masses of the galaxy particles range from $\sim10^{6}$ to
$\sim10^{9}M_{\odot }$; thus many galaxy particles are contained in a single
luminous galaxy in the real universe. Rather than group the particles into
galaxies, we simply use the galaxy particles themselves, which makes
the results less dependent on resolution.

Luminosities (in the relevant bands) are assigned to each cell following the
Bruzual and Charlot (1993, 1998; hereafter BC) model; we use their instantaneous
star-formation model, which best fits observations (Nagamine, Cen \&
Ostriker 1999). We also analyze our results using other BC models; the
main conclusions are insensitive to the specific star-formation model used.
The luminosities determined for each cell are summed over the galaxy
particles in the cell and evolve with time as given by the BC model.
The simulated luminosities are in excellent agreement with the observed
luminosity density in the universe at different redshifts
(Nagamine, Cen \& Ostriker 1999; see also below).

With the above information we can now determine the mass-to-light ratio, 
$M/L_{\rm B}$ (where $L_{\rm B}$ is the light in the blue band) at different
locations in the simulation volume and study it as a function of scale. In
order to minimize possible uncertainties due to model luminosities, we
normalize all luminosities - and thus $M/L_{\rm B}$ - to the \emph{observed}
luminosity density of the universe, as discussed below; this ensures that
our results are largely independent of the specific luminosity model used.

The behavior of $M/L_{\rm B}$ on small scales is determined in a similar
manner using smaller ($11.1h^{-1}$Mpc box), higher-resolution ($5h^{-1}$kpc)
Tree SPH simulations (see Dave \emph{et al.} 1999). This simulation uses a
similar cosmological model ($\Omega =0.4$, $\Omega_{\wedge }=0.6$,
$h=0.65$, $\sigma_{8}=0.8$); the small difference between the models is
adjusted in the final normalization of $M/L_{\rm B}$, but is insignificant. 
An $\Omega=1$ CDM model, tilted with $n=0.8$, is also investigated using this
simulation size. Galaxies are identified using SKID (see Katz, Weinberg \&
Hernquist  1996), and luminosities are assigned to each galaxy using the
same BC model described above.

\section{DEFINITION OF BIAS}

The term ``bias" has been used with different explicit
and implicit definitions, so it is essential that we be clear.
Kaiser (1984) introduced bias as the difference between the amplitude
of the correlation function of high density regions (such as galaxies
and clusters) relative to that of the mass in order to explain the
exceptionally strong correlation function observed for rich clusters
of galaxies (Bahcall \& Soneria 1983, Klypin \& Kopylov 1983). Similarly,
Davis et al. (1985) introduced bias as the proportionality constant
between the observed fluctuations in the number density of galaxies
and the mass fluctuations found in simulations:
$(\Delta N/N)_{\rm gal}\equiv b_{\rm gal} (\Delta \rho/\rho)_{m}$.
Since some smoothing scale ($R$) must be utilized to calculate
either side of the equation, bias must be explicitly a function
of scale, $b_{\rm gal}(R)$. Implicit were observational criteria
limiting the counted galaxies to be above a certain luminosity
and surface brightness.  If (and it is a substantial assumption)
one identifies the number density of halos in simulations with the number
density of galaxies then good dark matter simulations (e.g., Jenkins
\emph{et al.} 1998, Kravtsov \& Klypin 1999, Colin \emph{et al.} 1999),
which can compute $(\Delta N/N)_{\rm halo}=b_{\rm halo}(R)(\Delta\rho/
\rho)_{m}$, provide useful information and indicate low bias at large scales
($b_{\rm halo}\sim 1$), significant positive bias (b$_{\rm halo}>1$) at
intermediate scales, and antibias ($b_{\rm halo}<1$) on small ($R<5h^{-1}$Mpc)
scales due to merging of halos.

Hydrodynamic simulations which seek to identify the site of galaxy
formation and estimate the formation rate can compute the mass overdensity
in galaxies, $(\Delta \rho/\rho)_{\rm gal}$, although poor resolution limits
their ability to identify individual objects and to compute the galaxy number 
overdensity $(\Delta N/N)_{\rm gal}$. Recent papers by Katz \emph{et al}
(1999) and Cen \& Ostriker (2000) find significant positive bias on
intermediate scales. Blanton \emph{et al} (1999) discuss in detail the
physical origin of this bias, and its dependence on scale. Einasto et al.
(1999) discuss the physical origin of bias in terms of the fraction of mass
that exists in the voids. The ``semi-analytic" approach seeks to combine
in simplified form elements of the physical approach utilized in the
hydrodynamic modelling with the detailed resolution obtainable from pure
N-body work and has produced suggestive and most useful comparisons with
observations (Cole \emph{et al.} 1994, Kauffmann \emph{et al.} 1997). All
of the above work find positive (but small) bias on large scales so it is
important to understand the sense in which we will identify antibias in
this work.

The best way of comparing simulations to observations is neither through
$(\Delta N/N)_{\rm halo}$ nor $(\Delta \rho/\rho)_{\rm gal}$, but via 
$(\Delta j/j)_{\rm gal}$, where $j$ is the light emitted by the galaxies
in some band (here we use $j_{\rm B}$ in the blue band). We then compare
$(\Delta j_{\rm B}/j_{\rm B})_{\rm gal}$ with the same observed quantity
(after correction for obscuration).

Figure 1 shows in three panels (for top-hat smoothing scales
$1.5, 5, 10h^{-1}$Mpc) the average and the dispersion of 
$(\Delta \rho/\rho)_{\rm gal}$ and $(\Delta j_{\rm B}/j_{\rm B})_{\rm gal}$
versus the total mass overdensity $(\Delta \rho/\rho)_{m}$. Points above
the diagonal line are positively biased and those below the line are antibiased.
The $(\Delta \rho/\rho)_{\rm gal}$ curves are similar to those shown
in Cen \& Ostriker (1992, 2000) as well as Blanton \emph{et al} (1999)
and indicate positive bias on all scales in our dense regions
(approaching no bias in the highest density regions on these scales),
and negative bias (antibias) in underdense regions (i.e., little or no
galaxies in the `voids').

But we see that the light density $j_{\rm B}$ is antibiased, both relatively
and absolutely, in the highest density regions at 1.5, 5 and $10h^{-1}$Mpc
scales: $(\Delta j_{\rm B}/j_{\rm B})/(\Delta\rho/\rho)_{m}<1$.
The effect is small but real and easily understood. At low redshift
the highest density regions typically represent rich clusters and superclusters
(for large smoothing scales of $\sim 1$ to $10h^{-1}$Mpc); the stars and
galaxies in such regions tend to be old.  This well-known observational
fact is clearly seen in the simulations (Blanton \emph{et al} 1999; Cen
\& Ostriker 2000); after clusters form (at $z\simeq1-2$) the member galaxies
reside within a hot medium ($T=10^7-10^8$K) for which cooling is inefficient
and further star formation is inhibited.  In such old dense regions massive
young blue stars are rare, and the light diminishes sharply with increasing
time, especially in the blue. Our Bruzual-Charlot (BC) models, which
incorporate standard stellar evolution, thus show relatively low blue light
levels in the highest density regions at the present time; {\it the age effect}
overcomes the slight bias to bring the typical values of
$(\Delta j_{\rm B}/j_{\rm B})$ below $(\Delta \rho/\rho)_{m}$  
and yield a small {\it antibias in the highest density regions}.
The large amount of intracluster gas in these systems may also contribute 
to the antibias.

For observers who, in general, have no direct access to the ordinate in
Figure 1, $(\Delta\rho/\rho)_m$, it is interesting to consider the
ratio of the bias in the high density regions (rich clusters and 
superclusters) to the bias in the more normal regions where most 
galaxies live at moderate overdensities. Since these latter have a
significant positive bias, the {\it relative antibias} of high density 
regions as compared to low density regions is a factor of $\sim$2-4
(depending on scale).

\section{THE MASS-TO-LIGHT FUNCTION}

We now turn to the determination of the expected mass-to-light ratio of
galaxy systems as a function of scale by investigating $M/L_{\rm B}$ for
different size volumes in the simulations. In the $100h^{-1}$Mpc box, we
investigate volumes with radii ranging from $R \sim1h^{-1}$Mpc to $62h^{-1}$Mpc
(the volume-equivalent radius of the full box). For each volume of radius $R$
we determine the total mass $M$ and the total light $L_{\rm B}$ within the
volume, and hence $M/L_{\rm B}$. The volumes are centered on
randomly selected ``galaxies'' in the box, for proper comparison with 
observations (i.e., we center on random cells with total galaxy
particle mass exceeding 10$^{11}$ or 10$^{12}$ M$_{\odot }$; the results are
insensitive to the specific threshold).  For a given radius, a large number
of volumes are selected; these random volumes represent a wide range of mass
overdensities.  Rich clusters and superclusters of galaxies populate the
highest overdensity regions (at their respective scales), while loose groups
and other galactic systems correspond to regions of lower overdensities.

The first questions we ask are: How does $M/L_{\rm B}$ depend on scale and on
the local overdensity - does it flatten and become constant on large scales?
And, does it vary with overdensity (at a given scale)?

The results are presented in Figure 2, together with the observational data
discussed in \S 2. The immediately apparent result is that $(M/L)(R)$ increases
with scale on small scales and flattens on large scales, as seen in the 
observations. Each of the $(M/L_{\rm B})(R)$ curves for $R \geq 0.9h^{-1}$%
Mpc represents the simulation results for the mean of all volumes with
overdensity above a given threshold (at any given scale, as indicated in
Fig. 2). \ The highest overdensities are selected to correspond to
observed rich clusters of galaxies ($\Delta \rho /\rho \gtrsim $190 and $%
\gtrsim $ 250 at $R=1.5h^{-1}$Mpc, where $\Delta\rho/\rho$ is the total mass
overdensity; this corresponds approximately to richness
class $\gtrsim $ 0 and $\gtrsim $ 1 clusters; Abell 1958); these are shown
by the top solid and dashed curves. The lower overdensity regions are presented
by the dot-dashed curves; these are typical for loose groups of
galaxies at $R \sim1h^{-1}$Mpc. To illustrate the trend of $(M/L_{\rm B})(R)$
with overdensity, we scale the density thresholds with radius (from
$R = 1.5h^{-1}$Mpc) assuming a density profile f $\rho (r)\propto r^{-2.4}$,
as suggested by observations (e.g., Bahcall 1977, 1999, Peebles 1993, Carlberg
\emph{et al.} 1997); the results are similar for other reasonable extrapolations.

Voids, which contain little or no light (galaxies) but do contain some
mass, exhibit very large $M/L_{\rm B}$ ratios (e.g., Figure 1); their 
contribution is of course included in the total $(M/L_{\rm B})_{\rm box}$
and $\Omega$ values discussed below since these values refer to the entire 
amount of mass in the box.

The solid curve marked $\Omega $ = 0.37 represents the mean $M/L_{\rm B}$
function for $\Delta \rho /\rho (R \leq  1.5h^{-1}{\rm Mpc}) \geq \ 190$
for the $\Omega $ = 0.37 simulation (converted to h = 1 for comparison with
the data). \ The same solid line is then scaled up and down to $\Omega $ = 1
and $\Omega $ = 0.16 respectively (the latter, as shown below, is our
best-fit value), using linear scaling with $\Omega$, as expected (see below).
The entire set of $(M/L_{\rm B})(R)$ curves for different overdensities is
presented only once, for clarity, for $\Omega $ = 0.16.

The shape of the $(M/L_{\rm B})(R)$ function is nearly independent of
the specific model luminosities used; all models, including models with
different but observationally acceptable initial mass function (eg.,
Salpeter 1955, Miller \& Scalo 1979, Scalo 1986, for $\gtrsim0.1M_{\odot}$),
yield essentially the same function shape.  In order to be independent of
possible uncertainties also in the normalization of the model luminosities,
we normalize $L_{\rm B}$ of the entire simulation---and thus $M/L_{\rm B}$ of
the full box--- to the \emph{observed} luminosity density of the universe. 
The local luminosity density of the universe (in total B band luminosity, 
corrected for extinction) is observed to be $j_{\rm B}=(2\pm0.4)10^8hL_{\odot ({
\rm B})}{\rm Mpc}^{-3}$ (Efstathiou \emph{et al.} 1988, Lin \emph{et al.}
1996, Carlberg \emph{et al.} 1997, Ellis 1997, Small \emph{et al.} 1998 and
references therein). Since the mass density of the universe is
$\rho=3\Omega H_0^2/8\pi G=\Omega\rho_{\rm crit}=2.78\times10^{11}\Omega
h^2M_{\odot}{\rm Mpc}^{-3}$, the universal mass-to-light can be expressed as
$M/L_{\rm B}\equiv\rho/j_{\rm B}=(1400\pm280)\Omega h M_{\odot}/L_{\odot ({\rm B})}$, where $L_{\rm B}$ is the total, extinction corrected blue luminosity at
$z\simeq0$.  We normalize our simulation box to have the observed luminosity
density of the universe, $j_{\rm B}$, as listed above; the $M/L_{\rm B}$ of
the full box is thus fixed at $(M/L_{\rm B})_{\rm box}=518h$ (for
$\Omega=0.37$). Our results are therefore independent of
the absolute value of the simulated luminosities. In fact, the
direct simulation yields $M/L_{\rm B} =520h$ for the box, strongly supporting
the appropriateness of the luminosity model used. Similarly, for $\Omega $%
=1, $M/L_{\rm B}$ is normalized to be $M/L_{\rm B} =1400h$ ($\Omega $=1), as
required. 

On scales smaller than $0.9h^{-1}$Mpc, the smaller, higher-resolution
simulation is used (\S 3) to determine $(M/L_{\rm B})(R)$ from $R \sim20$
kpc to $\sim6h^{-1}$Mpc. Since the box is small, no high-density
regions such as rich clusters are found (since these are rare objects).
The $(M/L_{\rm B})(R)$ presented in Figure 2 represents the average of
typical bright galaxies (corresponding approximately to 
overdensities above the threshold indicated by the dot-dash curve, as
extrapolated to the smaller radii). The results are presented for $\Omega
= 0.16$ (scaled down from $\Omega=0.4$). The two sets of simulations agree
well with each other in the overlap region of $\sim$1 to $6h^{-1}$Mpc, thus
strongly supporting these independent results.

The results of Figure 2 show that the simulated $(M/L_{\rm B})(R)$ function
increases on small scales and then flattens on large scales as suggested by
observations (Bahcall \emph{et al.} 1995); the data and simulations exhibit
the same overall shape of the $(M/L_{\rm B})(R)$ function. This result is
independent of the specific luminosity model used; all models yield the
same basic $(M/L_{\rm B})(R)$ shape. Even though $M/L_{\rm B}$ flattens to a
constant value on large scales, a clear dependence of $(M/L_{\rm B})(R)$ on
the local overdensity (within a given radius R) is apparent; high overdensity
regions exhibit higher $M/L_{\rm B}$ ratios than lower density regions. The
results indicate that high density regions (such as rich clusters and
superclusters) are \emph{ antibiased} with respect to the mean, exhibiting
higher $M/L_{\rm B}$ ratios than average; this implies that mass is more
concentrated than light in the high density regions. This effect, as noted in
the previous section, is likely caused by the age effect: high density
clusters and superclusters are old systems, with low recent star-formation
(and thus lower than average blue luminosity); the old galaxies that dominate
these system have significantly reduced luminosities at this late time in
their evolution. Since all measures of $\Omega$ that utilize the $M/L_{\rm B}$
method use clusters and superclusters of galaxies --- which
are shown here to overestimate the mean $M/L_{\rm B}$ of the universe --- 
these measures also overestimate $\Omega$.

We can now determine an unbiased $\Omega $ by properly matching the simulated
$(M/L_{\rm B})(R)$ function to the data. As illustrated in Figure 2, both
$\Omega=1$ and $\Omega=0.37$ greatly overestimate the observed $M/L_{\rm B}$
ratio of groups, clusters, and superclusters, on all scales, by a factor of
$\sim 6$ (for $\Omega $ = 1) and $\sim$2 (for $\Omega $ = 0.37). \ This
overestimate is seen for the \emph{entire }observed range of the $M/L_{\rm B}$
function, not just for the classical case of clusters at $\sim1h^{-1}$Mpc. 
By fitting the entire observed and simulated mass-to-light
function - properly matching to the relevant overdensities - we can 
determine an unbiased measure of $\Omega ;$ we discuss this below (\S 6).

In Figure 3 we compare the observed $(M/L_{\rm B})(R)$ data with the simulated
results for the relevant high- and low- overdensity regions. The high
overdensity region (represented by the higher of the two bands at $R\gtrsim
1 h^{-1}$Mpc) corresponds to typical rich clusters and superclusters of
galaxies (at $\sim1.5h^{-1}$ and $5-20h^{-1}$Mpc respectively;
see specific overdensities listed in Figure 3). The low density region reflects
environments typical of looser groups and other galaxy systems. The
results are presented for both $\Omega $ = 0.16 and $\Omega $ = 1, as scaled
from the $\Omega $ = 0.37 simulation. \ On small scales, $R \simeq $ \ 20
kpc to $\sim6 h^{-1}$Mpc, the results from the high-resolution simulation
reflect the full $M/L_{\rm B}$ range obtained for individual galaxies
and groups. These results are in full agreement with the large-scale
simulations; the two independent results merge nicely with each other in the
overlap region. The $\Omega=1$ results on small scales
($\lesssim6h^{-1}$Mpc) are obtained directly from the $\Omega $ = 1 high 
resolution simulations; these direct simulation results
agree well with the scaled-up results from low $\Omega $ 
thus supporting the linear scaling of $M/L_{\rm B}$ with $\Omega $ on large
scales.

\section{DETERMINING $\Omega$}

The results presented in Fig. 3 provide a powerful illustration that an $\Omega
$ = 1 model significantly overestimates $M/L_{\rm B}$ on \emph{all scales}.
On large scales, the high overdensity band that represents typical rich
clusters (at $R \simeq  1.5h^{-1}$Mpc, $\Delta \rho /\rho \gtrsim 250$)
overestimates the observed $M/L_{\rm B}$ value for clusters by a factor of
$\sim$6--- a familiar result. A similar overestimate is seen for smaller
groups of galaxies, for individual galaxies, and for superclusters. Even an
$\Omega $ = 0.37 model appears to overestimate $(M/L_{\rm B})(R)$, by a factor
of $\sim$2 (with lower significance). 

To determine the best fit value of $\Omega $, we use two methods. In the
first method, we use the observed $M/L_{\rm B}$ ratio of rich clusters of
galaxies, and correct it to the proper global universal value (i.e., correct
for the cluster antibias) by using the simulation's ratio of $M/L_{\rm B}$
for the entire box to that of rich clusters. This ratio, $b_{\rm cl}^{^{M/L_{\rm B}}}\equiv[(M/L_{\rm B})_{\rm box}/<M/L_{\rm B}>_{\rm cl}]_{\rm sim}$,
is the bias factor (in $M/L_{\rm B}$) of clusters. For rich clusters (richness
class $\gtrsim $ 1) at $R \simeq 1.5h^{-1}$Mpc, we find 
\begin{equation}
b_{\rm cl}^{^{M/L_{\rm B}}}\equiv\left[\frac{<M/L_{\rm B}>_{\rm box}}{<M/L_{\rm B}>_{\rm cl}}\right]_{\rm sim}=0.75\pm0.15.\\
\end{equation}
The universal $M/L_{\rm B}$ value is thus given by $<M/L_{\rm B}>_{\rm cl}
\times b_{\rm cl}$; rich clusters overestimate the mean value by
a factor of $1/b_{\rm cl}\simeq 1.3$. The error-bar in (1) reflects the rms
scatter among the simulated cluster $M/L_{\rm B}$ values and the scatter
among the different luminosity models investigated. Since only the relative
ratio between the simulated $(M/L_{\rm B})_{\rm box}$ and $<M/L_{\rm B}>_{\rm
cl}$ is used in this method, the luminosity normalization is unimportant. The
mass density of the universe can be determined from the mean observed 
$M/L_{\rm B}$ of rich clusters (richness $\gtrsim $ 1) at $R\simeq 1.5h^{-1}$%
Mpc, $<M/L_{\rm B}>^{\rm obs}_{\rm cl} = 300 \pm 70h M _{\odot }/L_{\odot }$
(BLD; Carlberg \emph{et al.} 1997; with $L_{\rm B}$ in our standard system,
at $z=0$), and the observed luminosity density of the universe, $j_{\rm B}$,
\begin{equation}
\rho_m=<M/L_{\rm B}>_o\times j_{\rm B}=<M/L_{\rm B}>_{\rm cl}^{\rm obs}\times
b_{\rm cl}^{^{M/L_{\rm B}}}\times j_{\rm B}
\end{equation}
where $<M/L_{\rm B}>_o$ is the universal value. Therefore
\begin{equation}
\Omega\equiv\frac{\rho_m}{\rho_{\rm crit}}=\frac{<M/L_{\rm B}>_{\rm cl}^{\rm obs}\times b_{\rm cl}^{M/L_{\rm B}}}{(M/L_{\rm B})_{\rm crit}},
\end{equation}
where $(M/L_{\rm B})_{\rm crit}\equiv\rho_{\rm crit}/j_{\rm B}$ is the value
required for a critical density universe ($\Omega $ = 1; see \S 5). Recent
observations of the local galaxy luminosity function, corrected to the
standard system of luminosity used here, yield $j_{\rm B}=(2\pm0.4)\times10^8h
L_{\odot}{\rm Mpc}^{-3}$ and thus $(M/L_{\rm B})_{\rm crit}=1400\pm280h M_{\odot}/L_{\odot}$ (Lin \emph{et al.} 1996, Carlberg \emph{et al.} 1997, Ellis 1997,
Small \emph{et al.} 1998). The conservative error-bar used above reflects the
scatter among the different measurements as well as their uncertainties. We
thus find
\begin{equation}
\Omega= \frac{(300\pm70)(0.75\pm0.15)}{1400\pm280} = 0.16\pm0.06.
\end{equation}
The representative $M/L_{\rm B}$ value of the universe is $<M/L_{\rm B}>_o$
=225 $\pm $ 70, as given by the numerator of (4).

A second method of determining $\Omega $ is fitting the entire observed
$M/L_{\rm B}$ function of galaxies, groups, clusters, and superclusters
(MS0302) to the simulated function, for the relevant overdensities. Here we
use the high $\Delta \rho /\rho $ band (Fig. 3) for rich clusters, the lower
bound of this band for the MS0302 supercluster, and the low $\Delta
\rho /\rho $ band for groups (the upper sub-band is used since it best
matches the group overdensities). The small-scale $R < 1h^{-1}$Mpc band is
used for fitting the observed galaxies and small groups of galaxies (at
$\lesssim 0.5h^{-1}$Mpc). Fitting the observed to simulated $M/L_{\rm B}$
function has a single free parameter: $\Omega$; the best $\chi^2$ fit
yields $\Omega $ = 0.16 $\pm $ 0.02. Since the box normalization is fixed at
the observed value of $j_{\rm B}=(2\pm0.4)10^8h$, corresponding to
$(M/L_{\rm B})_{\rm box} = (1400\pm280)\Omega h$ (\S 5), the
result is essentially independent of the luminosity models. The result does
depend however on the observed normalization $j_{\rm B}$; therefore 
$\Omega=0.16\pm0.02(j_{\rm B}/(2\pm0.4)10^8h$), or equivalantly,
$\Omega =0.16\pm0.02[(1400\pm 280)h/(M/L_{\rm B})_{\rm crit}]$. Allowing for
the normalization uncertainty as well as for uncertainties in the overdensities
and in model luminosities, we find
\begin{equation}
\Omega =0.16\pm 0.05.
\end{equation}

This value is consistent with the one obtained earlier using clusters of
galaxies alone. Additional systematic uncertainties, while difficult to
accurately determine, may contribute an additional $\sim$ 20\%($\pm$ 0.03) to
the above uncertainty (see below). The $M/L_{\rm B}$ function for this
best-fit value, plotted in Fig. 3, reproduces well the entire observed 
$M/L_{\rm B}$ function, from galaxies to superclusters.


%
The error-bars given in (4,5) above may not include all possible systematic
uncertainties. For example, if low surface brightness galaxies contribute
significantly to the total luminosity density of the universe (over and above
the extrapolated luminosity function), but not to the luminosity in clusters,
this will increase $j_{\rm B}$ (thus decrease $(M/L_{\rm B})_{\rm crit}$) from
the value used, therefore increasing $\Omega $. However, if such galaxies
exist also in groups, clusters, and superclusters --- this effect will cancel
out. The effect, if exists, is expected to be small, and is at least partially
covered by the large uncertainty adopted for $j_{\rm B}$ and $(M/L_{\rm B})_{\rm crit}$. Similarly, a diffuse intracluster light, which may
account for $\sim$ 15\% of the total cluster luminosity (Feldmeier et al. 2000),
is not included in the observed cluster luminosity (it may in fact compensate
for the contribution of low surface brightness galaxies in the field). If
included, this will lower $<M/L_{\rm B}>^{\rm obs}_{\rm cl}$ 
and thus lower $\Omega $ (by $\sim$ 15\%). Systematic uncertainties in the
simulations may also contribute --- but only if they are scale dependent
(since the overall normalization is independent of the simulations).
It is unlikely that significant shape changes exist on the scales considered
here.  While difficult to accurately determine such possible systematic
uncertainties, we estimate that they may contribute an additional 
$\sim$ 20\% uncertainty to $\Omega $.


Figure 2 and 3 illustrate that the $(M/L_{\rm B})(R)$ function of high density 
regions increases with scale to an above-average peak at a
clusters-superclusters scale of few Mpc, then decreases to the mean universal
value. Conversely, low density regions reveal lower $M/L_{\rm B}$ values. 
This is consistent with observations of rich clusters (high density) versus
groups (low density); groups exhibit lower $M/L_{\rm B}$ ratios than typical
rich clusters by a factor of nearly two, as seen in both data and simulations.
Based on the present results we also expect that observations of weak lensing
in the ``field'', which are currently underway, will reveal lower $M/L_{\rm B}$
ratios than seen in clusters or superclusters of galaxies by a factor of up to
$\sim$2, depending on the specific overdensities.

Our best-fit $\Omega$ (eq. 5) is lower than previous estimates due to
the antibias discussed above as well as the more robust use of the entire $M/L$
function --- not just clusters ---in constraining $\Omega$.
A mass-density of $\Omega \simeq 0.35,$ frequently regarded as a current
''most popular'' value, appears to overestimate the entire observed $M/L$
function, \emph{on all scales}$,$ for galaxies, groups, clusters and
superclusters.

The above analysis uses overdensities selected in the
$\Omega =0.37$ simulation (keeping the same overdensities for the different
$\Omega$'s).
The actual overdensities (of groups, clusters, superclusters) in the lower
$\Omega\simeq0.16$ universe are of course twice as large, which can further
reduce $\Omega$ by $20\%$, to $\Omega\simeq0.13$. However, this effect,
which is caused mainly by the earlier cluster formation time in lower $\Omega$
models is minimized by the fact that {\it all} objects form earlier in such
models.
If so, the 
overdensities need not be re-scaled. If they are re-scaled, the best-fit
$\Omega$ may be lower than given above (by $\sim20$\%).

\section{ELLIPTICAL AND SPIRAL GALAXIES}

On small scales, the data show that $M/L_{\rm B}$ of elliptical galaxies is
larger than that of spirals by a factor of $\sim$3 (BLD; see also Tully and
Shaya 1998); this is mostly due to lower blue luminosity in the older
ellipticals, but could also be partially due to higher elliptical
mass. To test this observation in the simulations, we identify old
and young galaxies (thus mostly ellipticals and spirals respectively) by
selecting galactic systems based on their redshift of formation. For
example, in the large simulation box we define regions of ``old'' galaxies
as those where the total galactic particle mass formed at high redshift
(e.g., $z > 1.9$) exceeds that which formed at low redshift
(e.g., $z < 0.6$) by a factor of five. \ Thus regions dominated
by old galaxies satisfy: $M_{\rm gal}(z<0.6)/M_{\rm gal}(z >1.9) < 0.2$. 
Similarly, regions dominated by ``young'' galaxies satisfy $M_{\rm gal}(z<0.6)/M_{\rm gal}(z >1.9) > 0.2$. \ Varying the specific redshift cuts
and the fractional threshold (0.2, 0.4, 0.6) does not affect the final
results discussed below.

 In Figure 4 we present the $M/L_{\rm B}$ function for the old and young
galaxies as discussed above (for $R \geq  0.9 h^{-1}$Mpc; solid and dashed
curves). \ These curves are superimposed on the high and low overdensity
bands from Fig. 3. \ The results show a strong correlation: \ the old galaxy
$(M/L_{\rm B})(R)$ function traces remarkably well the high overdensity regions
(such as clusters and superclusters), while the young galaxies trace well
the low overdensity regions. \ No re-normalization has been applied, and the
results are insensitive to reasonable changes in the redshift and threshold
definitions of the young and old regions. \ This result is consistent with
observations in the sense that high density regions are indeed best traced
by old galaxies. \ The difference between $M/L_{\rm B}$ of the old and young
galactic regions is approximately a factor of 2 to 3, consistent with
observations. \ Extending the results to smaller scales of individual
galaxies, we select old and young galaxies in the high-resolution
simulations based on their colors: $B-V > 0.65$ (old) and $B-V
< 0.65$ (young). \ We plot in Fig. 4 the mean 10\% highest and
lowest $(M/L_{\rm B})(R)$ for galaxies in these respective color cuts, for 
$R \simeq 20$ kpc to 6 Mpc. \ The results depend only slightly on the exact
cuts. \ The results are consistent with those obtained from the large
simulation; they merge with each other in the overlap regions. The simulated
results are consistent with the data for the entire $(M/L_{\rm B})(R)$
function if - and only if - $\Omega $ $\simeq$0.16, as shown in Figs. 2-4.

\section{ CONCLUSIONS}
\qquad We use large-scale cosmological simulations to determine the expected
mass-to-light ratio of galaxy systems and its dependence on scale. 
The $(M/L_{\rm B})(R)$ function is investigated from small scales of galaxies
($R \simeq 20$ kpc) to large scales ($R \simeq 60h^{-1}$Mpc), and compared with
observations of galaxies, groups, clusters, and superclusters. We use the
results to evaluate the amount of bias on different scales (i.e., how mass
traces light), and use the comparison with observations to determine the
mass density of the universe, $\Omega $.

We find the following results:

\begin{enumerate}
\item In high density regions the galaxy blue light is antibiased (i.e., lower) relative to the total mass density (while the galaxy mass density is not). This is due to the old age of the high density systems which leads to a relative decrease in their
present-day luminosity, especially in the blue band that traces recent star formation.

\item The shape of the simulated $(M/L_{\rm B})(R)$ function is in excellent
agreement with observations.
The simulated $M/L_{\rm B}(R)$ function increases with
scale on small scales and flattens on large scales, where $M/L_{\rm B}$ reaches
a constant value, as observed.
The mean flattening of $(M/L_{\rm B})(R)$ on large scales indicates that, on
average, mass follows light on large scales (i.e., $M \propto L$).

\item Even though $M/L_{\rm B}$ is approximately constant on large scales, we
find that the actual value of $M/L_{\rm B}$ depends on the local mass
overdensity, $\Delta \rho /\rho (<R)$, at a given scale. \ High
overdensity regions exhibit higher $M/L_{\rm B}$ ratios than lower density
regions. \ The difference can typically be a factor of 2 to 3, consistent
with observations of groups and clusters of galaxies (representing low and
high density regions, respectively).
The dependence of $M/L_{\rm B}(R)$ on overdensity indicates that high
density regions such as rich clusters and superclusters are relatively \emph{antibiased%
} - they exhibit higher than average $M/L_{\rm B}$ values, implying that mass is
more concentrated than light in these regions (see 1 above). In the blue luminosity band, the
cluster $M/L_{\rm B}$ antibias is $b_{\rm cl}^{^{M/L_{\rm B}}}=<M/L_{\rm B}>_o/<M/L_{\rm B}>_{\rm cl}=0.75\pm0.15$.

\item We find that the $(M/L_{\rm B})(R)$ function of high density regions is
traced well by $(M/L_{\rm B})(R)$ of old (elliptical) galaxies; low density
regions are traced well by young (spiral) galaxies. \ These results are
consistent with observations.

\item We determine the mass density of the universe by fitting the simulated
$(M/L_{\rm B})(R)$ function to observations. 
The best fit $\Omega$ is lower than previous estimates based on cluster $M/L$
values because of the antibias discussed above as well as the more robust use 
of the entire $M/L$ function --- not just clusters --- in constraining $\Omega$.
We find a best-fit value of 
$\Omega=0.16\pm0.05$ (with an additional estimated uncertainty of $\pm0.03$ for possible additional systematics); this 
value provides a remarkably good match to the
data for galaxies, groups, clusters, and superclusters. \ The results are
independent of the details of the models and provide a
powerful measure of $\Omega$.
The only significant uncertainty we are aware of is due to the possibility
that current observations may systematically underestimate the global
mean luminosity density of the universe. This will produce a corresponding
underestimate in our computation of $\Omega$ unless there was also a 
corresponding underestimate in the luminosity of groups, clusters, and
superclusters of galaxies.
\end{enumerate}

\acknowledgments{
We thank J. Peebles, D. Spergel, P. Steinhardt, and M. Strauss for helpful
discussions. This work was supported by NSF grants AST-9803137 and ASC-9740300.
}

\newpage
\figcaption[fig1.ps]{
The galaxy mass overdensity, $\rho_{\rm gal}/<\rho_{\rm gal}>$ (=1+$(\Delta\rho/\rho)_{\rm gal}$), and the galaxy 
luminosity overdensity, $j_{\rm B}/<j_{\rm B}>$, are presented as a function of the total 
mass overdensity, $\rho_m/<\rho_m>$ (=1+$(\Delta\rho/\rho)_{m}$), for volumes with radii 1.5, 5, 
and $10h^{-1}$ Mpc (left to right panels) in the $100h^{-1}$ Mpc simulation. 
(The denominators represent the values of the full box.) 
Regions above the dotted diagonal line represent
positive bias ($b_{\rm gal}$ or $b_{j_{\rm B}}$ $>$1 ), while regions below the line are 
antibiased ($b_{\rm gal}$ or $b_{j_{\rm B}}$ $<$ 1); see \S 4.}

\figcaption[fig2.ps]{The mass-to-light function of galaxy systems from observations ($%
\S $2) and simulations ($\S $3,5). \ The data points include galaxies
(spirals and ellipticals, as indicated by the different symbols), groups,
rich clusters (at $R = 1.5h^{-1}$Mpc), supercluster (MS0302 at $\sim6h
^{-1}$Mpc, from weak lensing observations), and the Virgo infall and Least
Action analysis (shown on the largest scales) (from BLD; $\S $2). \ The
curves are the mean simulations results for regions above different
mass overdensity thresholds (listed above). \ The simulated $(M/L_{\rm B})(R)$ function
for $\Omega $ = 0.37 is scaled up and down to $\Omega $ = 1 and $\Omega $ =
0.16 respectively. \ Only the high-density solid curve is repeated for all
three models; the complete set of curves is shown for one case only ($\Omega 
$ = 0.16, our best-fit value). \ (The overdensities $\Delta \rho /\rho $
refer to the $\Omega $ = 0.37 simulation.) \ On small scales, the curve
represents the mean high-resolution simulation results for typical galaxies
and small groups.}

\figcaption[fig3.ps]{The mass-to-light function of galaxy systems from observations
and simulations. \ The simulated results (for $\Omega $ = 1 and 0.16, scaled
from $\Omega $ = 0.37) are presented for high- and low- density regions (for
$R \gtrsim 1h^{-1}$Mpc) typical of rich clusters/superclusters and small
groups of galaxies, respectively. \ The high-density region (top band)
represents the overdensities listed in the highest overdensity column (and
is bounded, bottom and top curves, by the mean 30\% lowest and 30\% highest
$M/L_{\rm B}$ values, respectively). The low-density band is outlined by
three $(M/L_{\rm B})(R)$ curves representing, bottom to top, the mean 
$(M/L_{\rm B})(R)$ for mass overdensities between the lowest
and next overdensity listed in the figure (from right to left; e.g., the
bottom curve is the mean for $\Delta \rho /\rho $ = 55 - 110 at $R = 0.9h
^{-1}$Mpc, etc.). \ The upper part of the band represents typical overdensities of small groups of
galaxies. \ On small scales, the wide band at $R \simeq $ 20kpc to 6Mpc
reflects $(M/L_{\rm B})(R)$ for the full range of typical galaxies and groups in
the high-resolution simulations (for $\Omega $ = 1 and $\Omega $ = 0.16). \
The, entire observed $M/L_{\rm B}$ function is well fit by the simulations only
if $\Omega \simeq 0.16.$}

\figcaption[fig4.ps]{The mass-to-light function from observations, simulations (for
high- and low- density regions; Fig. 3), and for simulated old and young
galaxies ($\sim$ ellipticals and spirals; \S 7). \ Old galaxies trace
well the high-density regions (with high $M/L_{\rm B}$ ratios) while young
galaxies trace well low-density regions (with lower $M/L_{\rm B}$ values). \ The
results are consistent with observations.}
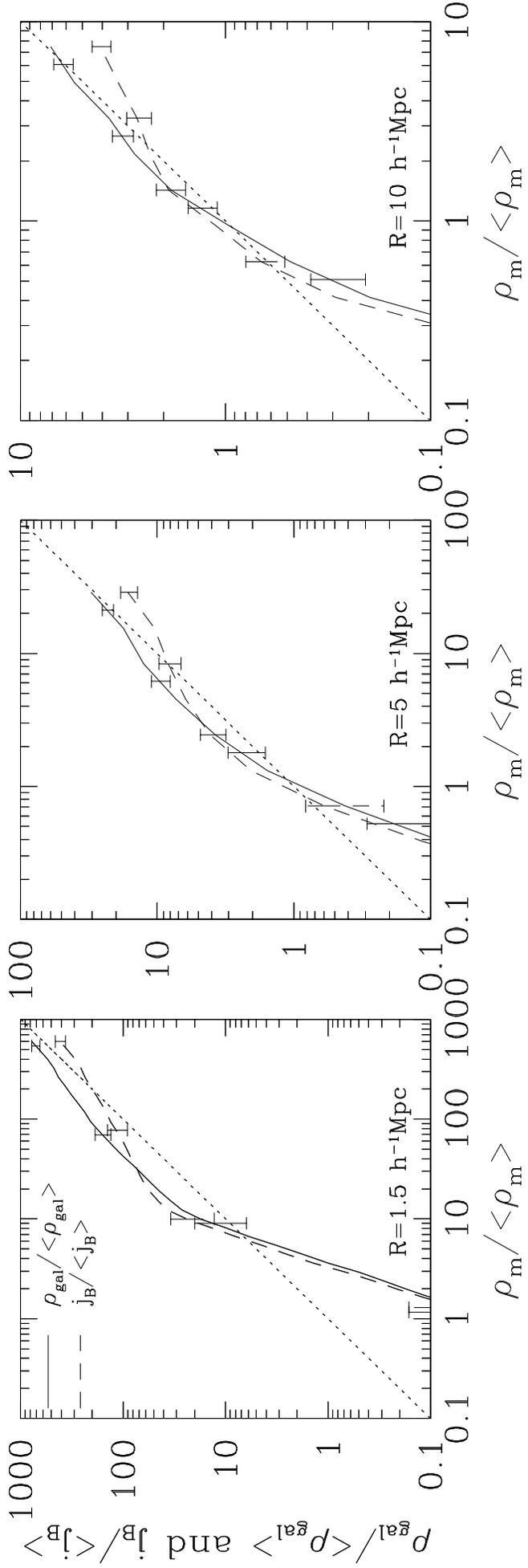
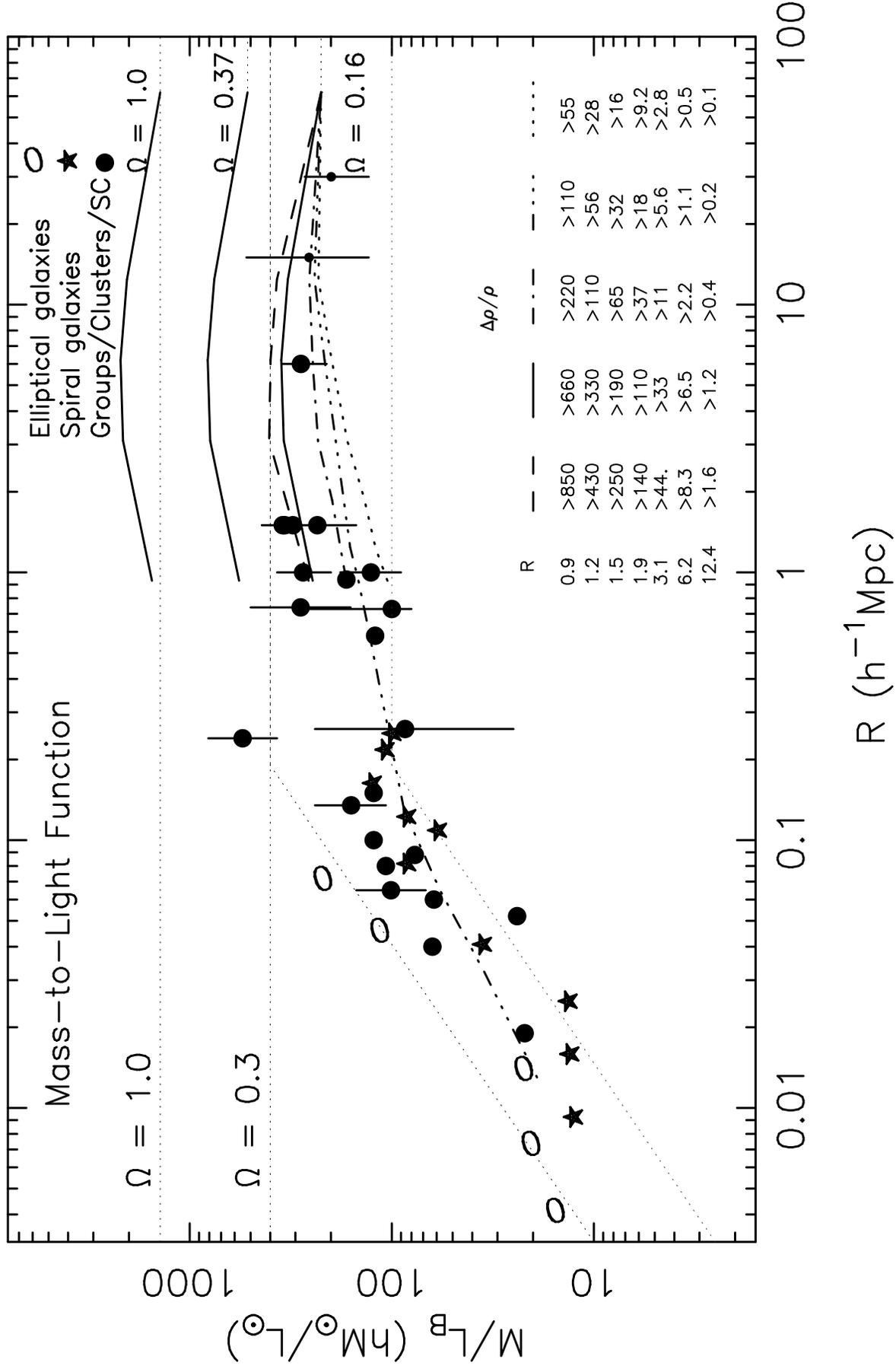
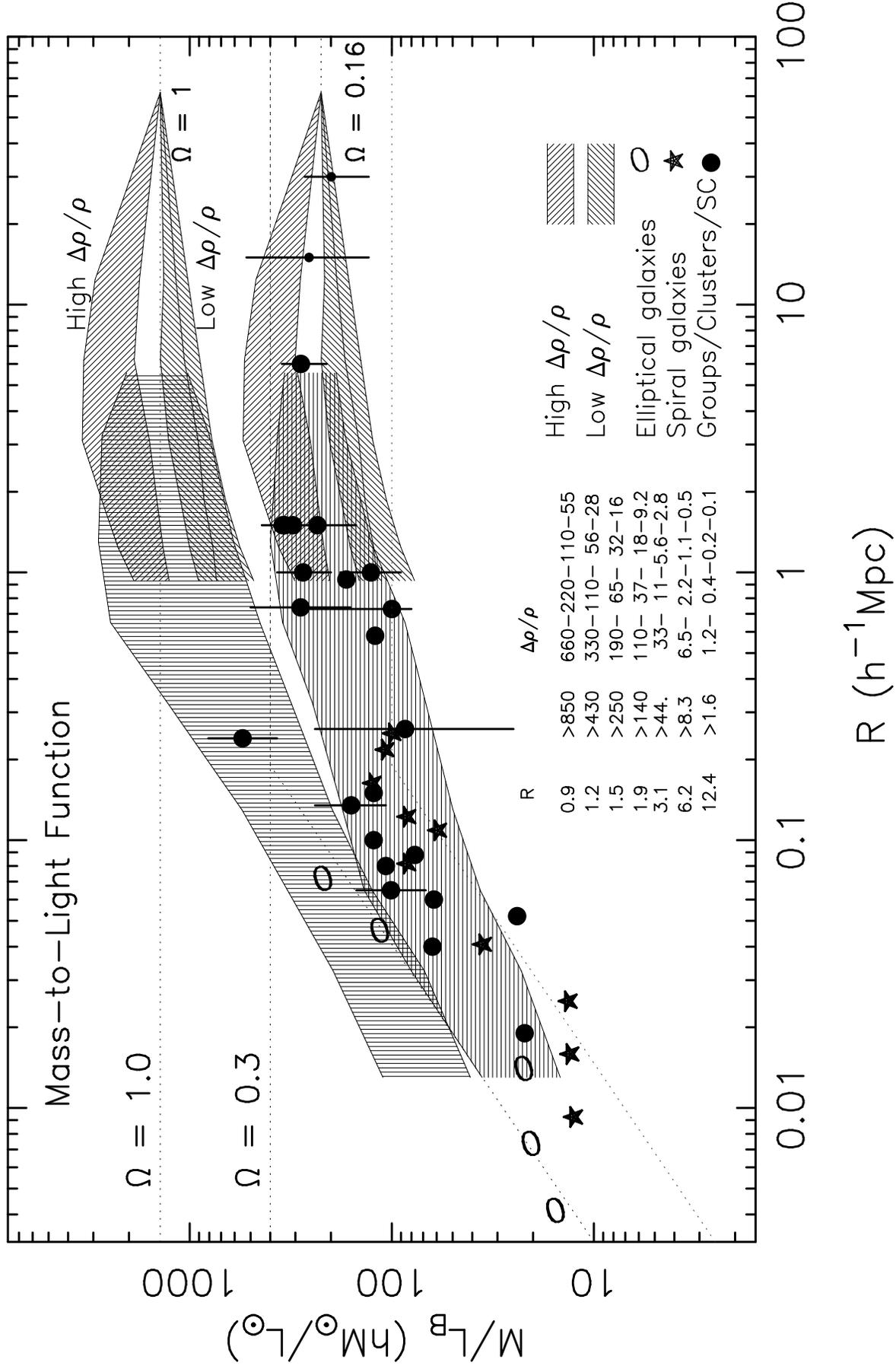
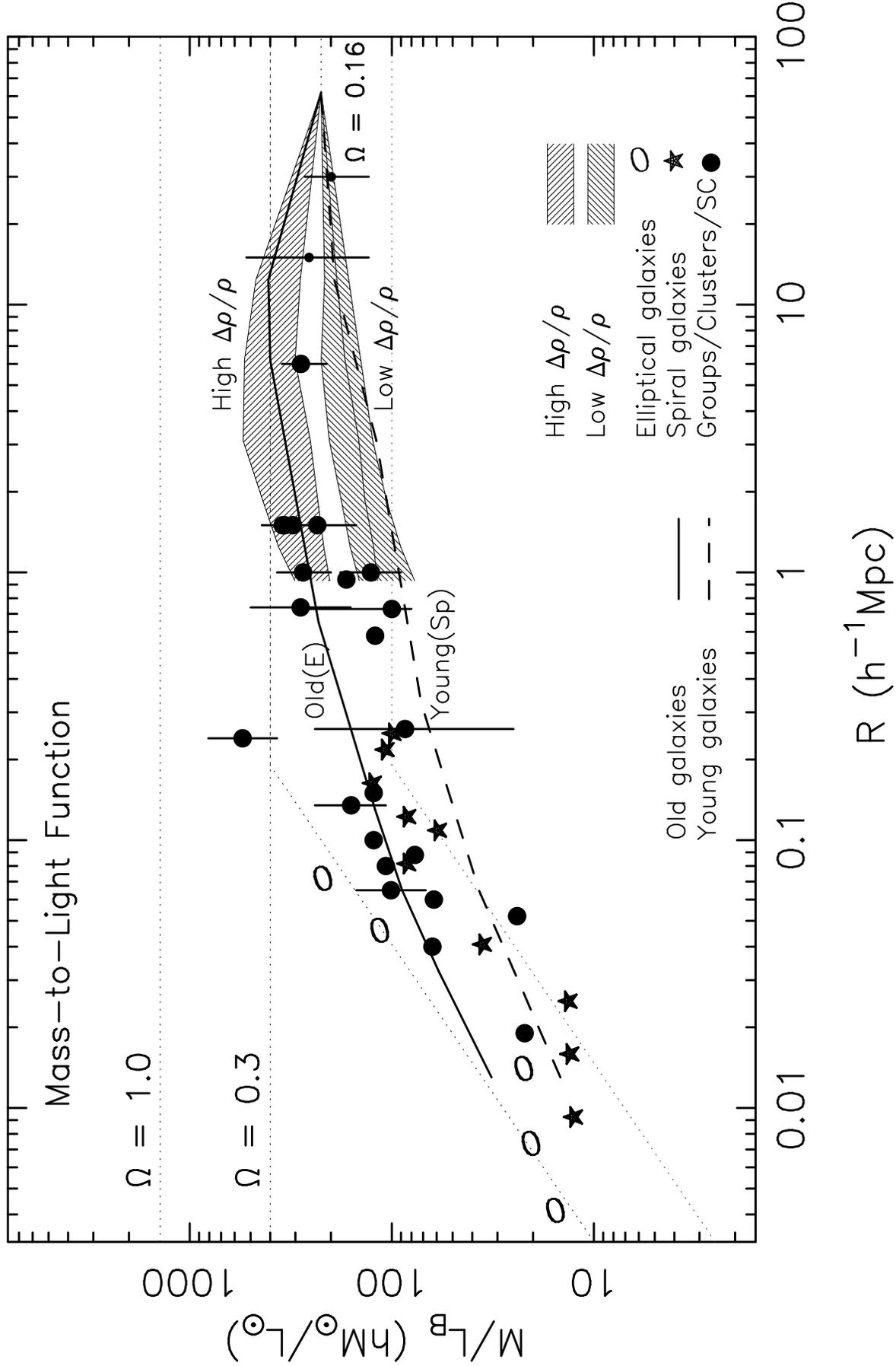
\end{document}